\documentstyle[12pt]{article}
\def\be{\begin{equation}}
\def\ee{\end{equation}}

\begin{document}

\title{The noncommutative replica approach}

\author{S.V.Kozyrev}
\maketitle

\begin{abstract}
$p$--Adic and non--commutative analysis are applied to describe
phase transitions in disordered systems. In the noncommutative
replica approach we replicate the disorder instead of the system
degrees of freedom. The noncommutatibe replica symmetry breaking
is formulated using the language of noncommutative analysis. This
allows to derive the ultrametric space of states which is
postulated in the standard replica approach.
\end{abstract}

\tableofcontents

\newpage

\section{Introduction}

In the present paper we apply $p$--adic and non--commutative
analysis to describe disordered systems. For introduction to
$p$--adic analysis see \cite{VVZ}, \cite{BS}. $p$--Adic analysis
and $p$--adic mathematical physics attract great interest, see
\cite{VVZ}--\cite{VV}. For instance, $p$--adic models in string
theory were introduced, see \cite{Vstring}, and $p$--adic quantum
mechanics \cite{VV} was investigated. $p$--Adic analysis was
applied to investigate the spontaneous breaking of the replica
symmetry, cf. \cite{cond-mat}, \cite{PaSu}. Similar approach was
used in \cite{Carlucci}, \cite{Carlucci1}.

We investigate disordered models such as the
Sherrington--Kirkpatrick model. Disorder in such models is
described by large stochastic matrices (the real symmetric
matrices with matrix elements being independent Gaussian
stochastic variables). The standard approaches to describe phase
transition in such systems are the replica approach and the Parisi
ultrametric anzats of breaking of replica symmetry
\cite{EA}--\cite{SpinGlass1}.

In the present paper we propose a new approach to describe phase
transitions in disordered systems based on non--commutative
analysis. We construct an analogue of the replica procedure in the
framework of non--commutative geometry and show that this
procedure naturally implies introduction of $p$--adic geometry.

To obtain our results we apply the following sequence of actions.

First, by the Wigner theorem the Sherrington--Kirkpatrick model in
the high temperature regime is described by the quantum Boltzmann
algebra in the free Fock representation. Note that the high
temperature regime here corresponds to the vacuum (zero
temperature) state of the quantum Boltzmann algebra. Then, the
transition from 1 to $p$ degrees of freedom is a morphism of
quantum probability spaces. We call this transformation the {\it
Non--commutative Replica Procedure}.

Second, we describe phase transition by putting the quantum
Boltzmann algebra into the state with non--zero expectation of
annihilators. To construct such a state we use the free coherent
states and the correspondent $p$--adic representation of the Cuntz
algebra. This representation acts on functions on the quantum line
which is equivalent to $p$--adic disk.

The free coherent states are eigenvectors of the linear
combination of annihilation operators from the quantum Boltzmann
algebra, or solutions $\Psi$ of the following equation of  {\it
the quantum line}
\begin{equation}\label{qline}
(A-\lambda)\Psi=0
\end{equation}
where $A$ is the linear combination of quantum Boltzmann
annihilators. In \cite{coherent2} it was shown that the space of
free coherent states is isomorphous to the state of generalized
functions on a $p$--adic disk. This means that {\it the quantum
line} (\ref{qline}) is equivalent to a $p$--adic disk (and $\Psi$
in (\ref{qline}) corresponds to a generalized function supported
at {\it the quantum line}).

Third, we conjecture that the $k$--th  correlator of the SK model
is equal to the expectation of the $k$--th degree of the order
parameter $Q=A+A^{\dag}$, $A={1\over\sqrt{p}}\sum_{i=0}^{p-1}A_i$
in the $p$--adic representation. This means that the corresponding
correlators will be given by $p$--adic integrals.

The structure of the paper is as follows.

In Section 2 we remind the Sherrington--Kirkpatrick model and
introduce the noncommutative replica procedure.

In Section 3 we put the Wigner theorem about the limits of large
stochastic matrices.

In Section 4 we discuss the quantum Boltzmann  and Cuntz algebras.

In Section 5 we describe the isomorphism of the free coherent
states and generalized functions on $p$--adic disk.

In Section 6 we discuss the $p$--adic representation of the Cuntz
algebra and its relation to the free coherent states.

In Section 7 we discuss the relation of phase transitions in
quantum systems and coherent states.

In Section 8 we formulate a conjecture describing the state of
disordered system (say of the SK model) after the phase
transition.

In Section 9 we discuss the results of the present paper.

\section{The SK model and replicas}

The typical disordered model is the SK (Sherrington--Kirkpatrick)
model with the Hamiltonian \be\label{SK} H=-\sum_{i<j}
J_{ij}\sigma^i \sigma^j \ee where the summation runs over the
spins $\sigma_{j}$ in the lattice $Z^d$ taking values $\pm 1$ and
$J_{ij}$ is the random matrix of interactions with the matrix
elements which are independent Gaussian random variables with the
probability distribution \be\label{J} P[J_{ij}]=\prod_{i<j}^N
\exp\left(-{NJ^2_{ij}\over 2}\right) \ee

Let us note that the stochastic matrix $J_{ij}$ is an example of a
quenched disorder: to calculate the expectation of an observable
we have to use some typical realization of the stochastic matrix
$J_{ij}$. To find the state one uses the assumption of
self--averaging: to calculate the expectation value of the
observable first we have to take the average over the spin degrees
of freedom and then take the average over the Gaussian stochastic
variables $J_{ij}$. The statistic sum takes the form
$$
Z=\int\sum_{\{\sigma_i\}} \exp\left(\beta \sum_{i<j}^N J_{ij}
\sigma_{i}\sigma_{j} -{N\over 2}\sum_{i<j}^N J^2_{ij}\right)
\prod_{i<j}^N  dJ_{ij}
$$

The well known approach to investigation of the
Sherrington--Kirkpatrick model and other disordered systems is the
replica method \cite{EA}--\cite{SpinGlass1}, where the spontaneous
breaking of the replica symmetry introduced by Parisi leads to
introduction of an ultrametric space. In the simplest case this
space coincides with a subset of the field $Q_p$ of $p$--adic
numbers and corresponding Parisi matrix is equal to the Vladimirov
operator of $p$--adic fractional derivation \cite{cond-mat},
\cite{PaSu}. In the replica approach one introduces $n$ identical
replicas of the original system. This implies the following
expression for the statistic sum \be\label{SKJ}
Z_n=\int\sum_{\{\sigma^a_i\}} \exp\left(\beta
\sum_{a=1}^n\sum_{i<j}^N J_{ij} \sigma^{a}_{i}\sigma^{a}_{j}
-{N\over 2}\sum_{i<j}^N J^2_{ij}\right) \prod_{i<j}^N  dJ_{ij} \ee
where $\beta$ is the inverse temperature. The transformation
$$
Z\mapsto Z_n
$$
is called the replica procedure \cite{EA}. The spontaneous
breaking of the replica symmetry \cite{Parisi} is the anzats
giving an approximation of this statistical sum.

In the present paper we develop a different approach to describe
phase transition in disordered system, based on noncommutative and
$p$--adic analysis. The starting point of this approach is to
replicate the disorder instead of the system degrees of freedom.

First, let us note that in the high temperature regime $\beta\to
0$, when we do not expect the phase transition, one can neglect
the spin part in the statistic sum and the Gibbs state of the
model will be given by (\ref{J}). In the thermodynamic limit
(\ref{J}) is described by the Wigner theorem, see the next
section.

Second, let us instead of replication of the system as in
(\ref{SKJ}), replicate the disorder. Consider $p$ independent
copies $J^{a}_{ij}$ of the random matrix $J_{ij}$ and modify the
statistic sum in the following way \be\label{Zp}
Z^{(p)}=\int\sum_{\{\sigma_i\}}  \exp\left({\beta\over\sqrt{p}}
\sum_{a=0}^{p-1}\sum_{i<j}^N J^{a}_{ij} \sigma_{i}\sigma_{j}
-{N\over 2}\sum_{a=0}^{p-1}\sum_{i<j}^N J^{(a)2}_{ij}\right)
\prod_{a=0}^{p-1}\prod_{i<j}^N dJ^{(a)}_{ij} \ee We call the
transformation
$$
Z\mapsto Z^{(p)}
$$
the {\it noncommutative replica procedure}. In this procedure
$p=n^{-1}$ and the well known $n\to 0$ limit of the replica
approach becomes $p\to\infty$ limit. Since in the high temperature
regime the state of the disordered system is described by the
Wigner theorem, for high temperatures the statistic sum above will
give an equivalent description of the SK model. In the
thermodynamic $N\to\infty$ limit, in the high temperature regime,
the state defined by the statistic sum (\ref{Zp}) will be
described by the noncommutative extension of the Wigner theorem
and will be give by the Fock state on the quantum Boltzmann
algebra
$$
A_iA_j^{\dag}=\delta_{ij}
$$
This observation explains the word noncommutative (replica
procedure).

For lower temperatures we will obtain the phase transition in the
described system, after which the state on the quantum Boltzmann
algebra will become non--Fock. In the present paper we introduce
the corresponding state and show, using the results of
\cite{coherent1}--\cite{coherent3}, that the introduced state is
related to $p$--adic geometry, which is the simplest example of
the ultrametric geometry. This observation is in agreement with
the results obtained in the replica symmetry breaking approach
\cite{SpinGlass}.

\section{Large random matrices and the Wigner theorem}

In order to obtain the deeper insight, let us discuss in details
the high temperature limit, in which the SK model is exactly
solvable.

Let $X=(X_{ij})_{i,j=1}^N$ be an ensemble of real symmetric
$N\times N$ matrices with distribution given by \be\label{1.10}
\langle f(X)\rangle_N={1\over Z_N}\int \hbox{tr }f\left({X\over
N}\right) e^{-{1\over 2}\hbox{tr } X^2}\prod_{i\leq j}dX_{ij}, \ee
$$
Z_N=\int e^{-{1\over 2}\hbox{tr } X^2}\prod_{i\leq j}dX_{ij}
$$
The limits \be\label{1.12} \lim_{N\to\infty}\langle X^k\rangle_N=
(\Omega,Q^k\Omega)\qquad;\qquad k=1,2,\dots \ee exist. The
limiting object $Q$ in (\ref{1.12}) is a quantum random variable
which is called the master field. It is defined as \be\label{1.13}
Q=A^{\dag} +A \ee where $A^{\dag}$ and $A$ are free creation and
annihilation operators, i.e. they do not satisfy the usual
canonical commutation relations but the following \be\label{1.14}
AA^{\dag} =1 \ee The vector $\Omega$ in (\ref{1.12}) is a vacuum
vector in the free Fock space,
$$
A\Omega =0
$$
The free Fock space is constructed starting from $A^{\dag}$ and
$\Omega$, by the usual procedure but only by using the operators
$A$ and $A^{\dag}$ satisfying (\ref{1.14}).

The expectation value in (\ref{1.12}) is known to be
\be\label{1.15} (\Omega,Q^k \Omega)=\int_R \lambda^k
w(\lambda)d\lambda \ee where $w(\lambda)$ is the Wigner semicircle
density, \be\label{1.16} w(\lambda)={1\over 2\pi}\sqrt
{4-\lambda^2}\qquad, \ \ |\lambda|\leq 2 \ee and $w(\lambda)=0$
for $|\lambda|\geq 2$. The Wigner semicircle distribution in
noncommutative probability plays the role of the Gaussian
distribution in classical probability.

This form of the Wigner theorem was presented in \cite{ALV}.
Discuss now a generalization of the Wigner theorem.

Consider the space $P(N,{\bf R})$ of polynomials of symmetric
$N\times N$ matrices over real numbers. Consider the state over
$P(N,{\bf R})$ (the same as in the Wigner theorem):
$$
\langle f(X)\rangle_N={1\over Z_N}\int\hbox{ tr }
f\left({X\over{N}}\right) e^{-{1\over 2}\hbox{tr }
X^2}\prod_{i\leq j}dX_{ij}
$$
Introduce $p$ independent copies of the space $P(N,{\bf R})$ and
the tensor degree $P(N,{\bf R})^{\otimes p}$. Introduce the state
on $P(N,{\bf R})^{\otimes p}$ in the following way
\be\label{expstate} \langle
f\left(X^{(1)},\dots,X^{(p)}\right)\rangle_N={1\over Z_N}\int
\hbox{ tr }f\left({X^{(1)}\over{N}},\dots,{X^{(p)}\over{N}}\right)
e^{-{1\over 2}\sum_{k=1}^{p}\hbox{tr }
X_k^2}\prod_{k=1}^p\prod_{i\leq j}dX^{(k)}_{ij} \ee
$$
Z_N=\langle 1\rangle_N
$$
The following result was presented in \cite{Voi92} and was used un
the $N\to\infty$ limit of the matrix model, see
\cite{AV}--\cite{ALV}.

\bigskip

\noindent{\bf Theorem 1}\qquad {\sl The limits (\ref{expstate})
exist for each polynomial $f$ and are equal to
$$
\lim_{N\to\infty}\langle
f\left(X^{(1)},\dots,X^{(p)}\right)\rangle_N=(\Omega,f(Q_1,\dots,Q_p)\Omega)
$$
where
$$
Q_i=A_i+A_i^{\dag}
$$
and $A_i$ are the quantum Boltzmann annihilators, satisfying
$$
A_iA^{\dag}_j=\delta_{ij}
$$
and $\Omega$ is the vacuum in the free Fock space:
$$
A_i\Omega=0
$$
}

\bigskip

\noindent{\it Idea of the Proof}\qquad  Discuss first the sketch
of the proof of the Wigner theorem.

The integral in the Wigner theorem is
$$
\langle f(X)\rangle_N={1\over Z_N}\int\hbox{ tr }
f\left({X\over{N}}\right) e^{-{1\over 2}\hbox{tr }
X^2}\prod_{i\leq j}dX_{ij}
$$

The following integral is equal to the Fock expectation of the
degree of the sum of Bose creator and annihilator: \be\label{bose}
{1\over\sqrt{2\pi}}\int x^k e^{-{1\over
2}x^2}dx=\langle(a+a^{\dag})^k\rangle;\qquad [a,a^{\dag}]=1,
\qquad\langle a\rangle=0 \ee For the matrix $X=(X_{ij})$ one get
\be\label{finiteN} \langle X^k\rangle_N={1\over Z_N}\int\hbox{ tr
} \left({X\over{N}}\right)^k e^{-{1\over 2}\hbox{tr }
X^2}\prod_{i\leq j}dX_{ij}= {N^{-{k}}\over Z_N}\int
X_{i_1i_2}X_{i_2i_3}\dots X_{i_{k}i_1} e^{-{1\over 2}\hbox{tr }
X^2}\prod_{i\leq j}dX_{ij} \ee where we assume the summation over
$i_j$.

Using (\ref{bose}) and the Wick theorem, we see that the
half--planar diagrams gives contribution of order $N^{k}$ into the
integral above, and the crossing diagrams give contribution of
lower order, see \cite{ALV} for details. In the large $N$ limit
only half--planar diagrams survive.

Since the limits of the correlation functions coincide with the
correlations on the quantum Boltzmann algebra, this implies that
the $N\to \infty$ limits of the correlators are reproduced by the
Fock state on the quantum Boltzmann algebra:
$$
\lim_{N\to\infty}\langle X^k\rangle_N=\langle
(A+A^{\dag})^k\rangle,\qquad AA^{\dag}=1,\qquad\langle A\rangle=0
$$

\bigskip

In the statement of theorem 1 we have the polynomials which are
linear combinations of the monomials of the form
$$
f(X^{(1)},\dots,X^{(p)})=X^{(j_1)}\dots X^{(j_k)}
$$
We have the state on the monomial
$$
\langle X^{(j_1)}\dots X^{(j_k)}\rangle_N={N^{-k}\over Z_N}\int
X_{i_1i_2}^{(j_1)}\dots X_{i_ki_1}^{(j_k)} e^{-{1\over
2}\sum_{k=1}^{p}\hbox{tr } X^{(k)2}}\prod_{k=1}^p\prod_{i\leq
j}dX^{(k)}_{ij}
$$
Using the same arguments which were used to prove the Wigner
theorem, we obtain the proof of theorem 1.

\bigskip

Consider the following linear map
$$
\Delta: P(N,{\bf R})\longrightarrow P(N,{\bf R})^{\otimes p}
$$
\be\label{onX} X\mapsto {1\over\sqrt{p}}\left(X\otimes 1\otimes
1\otimes\dots\otimes 1+1\otimes X\otimes 1\otimes\dots\otimes
1+\dots+ 1\otimes 1\otimes \dots\otimes 1\otimes X\right) \ee or
equivalently
$$
X\mapsto {1\over\sqrt{p}}\sum_{i=1}^{p} X_i
$$
where $X_i$ belongs to the $i$--th component of the tensor
product. This map coincides with the map used in the central limit
theorem. For example, for the gaussian random variables $X$ the
map (\ref{onX}) is an embedding of probability spaces (all the
correlation functions are invariant). For large random matrices,
in the thermodynamic (large $N$) limit, the central limit theorem
becomes the free central limit theorem, see \cite{Voi92}. For
instance, the Wigner state will be invariant under (\ref{onX}). We
formulate the following:

\bigskip

\noindent{\bf Theorem 2}\qquad {\sl The map (\ref{onX}) with the
state  (\ref{expstate}), where we put
$$
f\left(X^{(1)},\dots,X^{(p)}\right)= f\left(\Delta X\right)
$$
is an embedding of algebraic probability spaces. In the limit
$N\to\infty$ this embedding becomes the $*$--homomorphism of the
quantum Boltzmann algebra with one degree of freedom maps into the
quantum Boltzmann algebra with $p$ degrees of freedom defined
according to the formula \be\label{algebra}
A\mapsto{1\over\sqrt{p}}\sum_{i=1}^{p} A_i,\qquad
A^{\dag}\mapsto{1\over\sqrt{p}}\sum_{i=1}^{p} A^{\dag}_i \ee and
the Fock state $\langle\Omega,x\Omega\rangle$ will map onto the
Fock state $\langle\Omega,x\Omega\rangle_p$ in the quantum
Boltzmann Fock space with $p$ degrees of freedom \be\label{state}
\langle\Omega,x\Omega\rangle\mapsto \langle\Omega,x\Omega\rangle_p
\ee The map (\ref{algebra}), (\ref{state}) conserves all the
correlators.}

\bigskip

\noindent{\it Proof}\qquad Follows from the Wick theorem and the
analysis of the correlators.

\bigskip

The substitute (\ref{onX}) we call the noncommutative replica
procedure for high temperature.

The embedding of the quantum probability spaces we understand in
the following way. The morphism of quantum probability spaces
$(A,\phi)$ and $(B,\psi)$, see \cite{ALV}, is the morphism of
algebras
$$
f:A\to B
$$
such that the correlators are invariant:
$$
\psi(f(a))=\phi(a)
$$
We call the morphism $f$ an embedding, if it is an embedding as
the map of algebras.

Summing up, we see that the Wigner theorem allows to compute the
thermodynamic limit of the SK model in the high temperature
regime. We introduce the noncommutative replica procedure
(\ref{onX}) and show, that in the high temperature regime it
realizes an equivalent description of the SK model (by theorems 1
and 2).

In the present paper we investigate phase transitions in the SK
model. Before the phase transition the noncommutative replica
procedure (\ref{onX}) is embedding of quantum probability spaces.
After the phase transition the noncommutative replica procedure
will give rise to non trivial state of the quantum Boltzmann
algebra, which we will approximate using the free coherent states.

\section{The quantum Boltzmann and Cuntz algebras}

In the present section we discuss the quantum Boltzmann and the
Cuntz algebras.

The quantum Boltzmann algebra arises in the limit of large
stochastic matrices \cite{Wig}, was used in the free probability
\cite{Voi92} and describes the quantum system in the stochastic
approximation \cite{book}--\cite{qdeform}. It is defined as
follows. The quantum Boltzmann, or free commutation relation
algebra (or FCR-algebra) over the Hilbert space  $H$ is the unital
involutive algebra with generators (which are called free, or
quantum Boltzmann, creation and annihilation operators)
$A^{\dag}(f)$, $A(f)$, $f\in H$, and relation
$$
A(f)A^{\dag}(g)=(f,g).
$$
The FCR-algebra have a natural representation in the free Fock
space. Free  (or Boltzmannian) Fock space ${\cal F}$ over a
Hilbert space $H$ is the completion of the tensor algebra
$$F=\oplus_{n=0}^{\infty}H^{\otimes n}.$$
Creation and annihilation operators are defined in the following
way:
$$
A^{\dag}(f) f_{1}\otimes \dots \otimes f_{n}=f\otimes f_{1}\otimes
\dots \otimes f_{n}
$$
$$
A(f) f_{1}\otimes \dots \otimes f_{n}=\langle f,f_{1} \rangle
f_{2}\otimes \dots  \otimes f_{n}
$$
where  $\langle f,g \rangle $ is the scalar product in the Hilbert
space $H$. Scalar product  in the free Fock space is defined by
the standard construction of the direct sum of tensor products of
Euclidean spaces.

In the case when  $H$ is the $p$-dimensional complex Euclidean
space we have $p$ creation operators $A^{\dag}_{i}$, $i=0,\dots
,p-1$; $p$ annihilation operators $A_{i}$, $i=0,\dots ,p-1$ with
the relations
\begin{equation}\label{aac}
A_{i}A_{j}^{\dag}=\delta_{ij}.
\end{equation}
and the vacuum vector $\Omega$  in the free Fock space satisfies
\begin{equation}\label{vacuum}
A_{i}\Omega=0.
\end{equation}
Let us consider an orthonormal basis in the free Fock space of the
form
$$
e_I= A^{\dag}_{I}\Omega;
$$
here  the multiindex $I=i_0 \dots i_{k-1}$, $i_j \in\{0, \dots
,p-1\}$ and $A^{\dag}_{I}=A^{\dag}_{i_{k-1}} \dots
A^{\dag}_{i_0}$.

In the considered basis the Fock representation of FCR-algebra has
the form
$$
A^{\dag}_{i}e_I= e_{Ii}.
$$
One also defines (in the same Hilbert space) the  Antifock
representation of FCR-algebra
$$
A^{\dag}_{i}e_I= e_{iI}.
$$
Fock and  Antifock  representation of FCR-algebra are unitarily
equivalent.

We will also use the following factor--algebra of the quantum
Boltzmann algebra. The Cuntz algebra (with $p$ degrees of freedom)
is the algebra with involution which is generated by $p$ creation
operators $A^{\dag}_{i}$, $i=0,\dots ,p-1$; $p$ annihilation
operators $A_{i}$, $i=0,\dots ,p-1$ with the relations
\begin{equation}\label{aac1}
A_{i}A_{j}^{\dag}=\delta_{ij};
\end{equation}
\begin{equation}\label{cuntz}
\sum_{i=0}^{p-1}A_{i}^{\dag}A_{i}=1.
\end{equation}

\section{The free coherent states}

Since the free coherent states \cite{coherent1}, \cite{coherent2}
are related to $p$--adic numbers, let us make here a brief review
of $p$--adic analysis. The field $Q_p$ of $p$--adic numbers is the
completion of the field of rational numbers  $Q$ with respect to
the $p$--adic norm on $Q$. This norm is defined in the following
way. An arbitrary rational number $x$ can be written in the form
$x=p^{\gamma}\frac{m}{n}$ with $m$ and $n$ not divisible by $p$.
The $p$--adic norm of the rational number
$x=p^{\gamma}\frac{m}{n}$ is equal to $|x|_p=p^{-\gamma}$.

The most interesting property of the field of   $p$-adic numbers
is ultrametricity. This means that $Q_p$ obeys the strong triangle
inequality
$$
|x+y|_p \le \max (|x|_p,|y|_p).
$$
We will consider disks in   $Q_p$ of the form $\{x\in Q_p:
|x-x_0|_p\le p^{-k}\}$. For example, the ring $Z_p$ of integer
$p$--adic numbers is the disk $\{x\in Q_p: |x|_p\le 1\}$ which is
the completion of integers with the $p$--adic norm. The main
properties of disks in arbitrary ultrametric space are the
following:

{\bf 1.}\qquad Every point of a disk is the center of this disk.

{\bf 2.}\qquad Two disks either do not intersect or one of these
disks contains  the other.

\bigskip

Discuss in short the results of \cite{coherent1},
\cite{coherent2}, \cite{coherent3}.

The free coherent states (or shortly FCS) were introduced in
\cite{coherent1}, \cite{coherent2} as the formal eigenvectors of
the annihilation operator
$A={1\over\sqrt{p}}\sum_{i=0}^{p-1}A_{i}$ in the free Fock space
${\cal F}$ for some eigenvalue ${\lambda\over\sqrt{p}}$,
\begin{equation}\label{freecoherent}
A \Psi= {\lambda\over\sqrt{p}} \Psi.
\end{equation}
The formal solution of (\ref{freecoherent}) is
\begin{equation}\label{psi}
\Psi=\sum_{I} \lambda^{|I|} \Psi_I A^{\dag}_{I}\Omega.
\end{equation}
Here  the multiindex $I=i_0 \dots i_{k-1}$, $i_j \in\{0, \dots
,p-1\}$ and \be\label{Adag} A^{\dag}_{I}=A^{\dag}_{i_{k-1}} \dots
A^{\dag}_{i_0} \ee $\Psi_I$ are complex numbers which satisfy
\begin{equation}\label{cascade}
\Psi_I=\sum_{i=0}^{p-1}\Psi_{Ii}.
\end{equation}
The summation in the formula  (\ref{psi}) runs on all sequences
$I$ with finite length. The length of the sequence $I$ is denoted
by $|I|$ (for instance in the formula above $|I|=k$). The formal
series (\ref{psi}) defines the functional with a dense domain in
the free Fock space. For instance the domain of each free coherent
state for $\lambda\in (0,\sqrt{p})$ contains the dense space $X$
introduced below.

We define the free coherent state $X_I$ of the form
\begin{equation}\label{indicator}
X_I= \sum_{k=0}^{\infty} \lambda^k \left(\frac{1}{p}
\sum_{i=0}^{p-1}A_i^{\dag}\right)^k \lambda^{|I|} A^{\dag}_I
\Omega+ \sum_{l=1}^{\infty} \lambda^{-l}
\left(\sum_{i=0}^{p-1}A_i\right)^l \lambda^{|I|} A^{\dag}_I \Omega
\end{equation}
The sum on $l$ in fact contains $|I|$ terms. For $\lambda\in
(0,\sqrt{p})$ the coherent state $X_I$ lies in the Hilbert space
(the correspondent functional is bounded).

We denote by $X$ the linear span of free coherent states of the
form (\ref{indicator}) and by $X'$ we denote the space of all the
free coherent states (given by (\ref{psi})).

The following definitions and theorems were proposed in
\cite{coherent1}, \cite{coherent2}, \cite{coherent3}.

\bigskip

\noindent{\bf Definition 3}\qquad {\sl We define the renormalized
pairing of the spaces $X$ and $X'$ as follows:
\be\label{renormalizedproduct}
(\Psi,\Phi)=\lim_{\lambda\to\sqrt{p}-0}\left(1-\frac{\lambda^2}{p}\right)
\langle \Psi,\Phi\rangle \ee Here $\Psi\in X'$, $\Phi\in X$. }

\bigskip

Note that the coherent states $\Psi$, $\Phi$ defined by
(\ref{psi}), (\ref{indicator}) depend on $\lambda$ and the product
$(\Psi,\Phi)$ does not.

\bigskip

\noindent{\bf Definition 4}\qquad {\sl We denote $\tilde{\cal F}$
the completion of the space $X$ of the free coherent states with
respect to the norm defined by the renormalized scalar product. }

\bigskip

The space $\tilde{\cal F}$ is a Hilbert space with respect to the
renormalized scalar product.

\bigskip

\noindent {\bf Theorem 5}\qquad {\sl The space of the free
coherent states \be\label{fcstriple} X
\stackrel{i}{\longrightarrow} \tilde{\cal F}
\stackrel{j}{\longrightarrow}  X' \ee is a rigged Hilbert space. }

\bigskip

Define the characteristic functions of $p$--adic disks
\begin{equation}\label{theta}
\theta_k(x-x_0)=\theta(p^{k}|x-x_0|_p);\quad \theta(t)=0,
t>1;\quad \theta(t)=1, t\le 1.
\end{equation}
Here $x$, $x_0\in Z_p$ lie in the ring of integer $p$--adic
numbers and the function $\theta_k(x-x_0)$ equals to 1 on the disk
$D(x_0,p^{-k})$ of radius $p^{-k}$ with the center in $x_0$ and
equals to 0 outside this disk.

We compare the rigged Hilbert spaces of the free coherent states
(\ref{fcstriple}) and of generalized functions over $p$--adic
disk:
$$
D(Z_p)\stackrel{i'}{\longrightarrow}
L^2(Z_p)\stackrel{j'}{\longrightarrow} D'(Z_p)
$$

\bigskip

\noindent {\bf Theorem 6}\qquad {\sl The map $\phi$ defined by
$$
\phi:\quad X_I\mapsto p^{|I|}\theta_{|I|}(x-I);
$$
extends to an isomorphism $\phi$ of the rigged Hilbert spaces:
$$
\begin{array}{ccccc}
X & \stackrel{i}{\longrightarrow} & \tilde{\cal F} &
\stackrel{j}{\longrightarrow} & X' \\
\downarrow\lefteqn{\phi}&&\downarrow\lefteqn{\tilde\phi}&&
\downarrow\lefteqn{\phi'}\\
D(Z_p) & \stackrel{i'}{\longrightarrow} & L^2(Z_p) &
\stackrel{j'}{\longrightarrow} & D'(Z_p)
\end{array}
$$
between the rigged Hilbert space of the free coherent states (with
the pairing given by the renormalized scalar product) and the
rigged Hilbert space of generalized functions over $p$--adic disk.

}

\bigskip

Here $I$ are $p$--adic numbers corresponding to multiindices: for
multiindex $I=i_0 \dots i_{k-1}$ corresponding $p$--adic number
(which we denote by the same symbol) equals to
$$
I=\sum_{j=0}^{k-1}i_j p^j
$$

Definition (\ref{freecoherent}) of the space of FCS:
$$
\left(A-{\lambda\over\sqrt{p}}\right)\Psi=0
$$
may be interpreted as the equation of the noncommutative (or
quantum) plane $A={\lambda\over\sqrt{p}}$. The free coherent state
$\Psi$ in this picture corresponds to a generalized function on a
non--commutative space (with non--commutative coordinates $A_i$,
$A^{\dag}_i$) with support on the non--commutative plane
$A={\lambda\over\sqrt{p}}$, for
$A={1\over\sqrt{p}}\sum_{i=0}^{p-1} A_i$.

The theorem above means that the space of generalized functions
over the non--commutative plane is isomorphic as a rigged Hilbert
space to the space of generalized functions over a $p$--adic disk,
or roughly speaking the non--commutative plane is equivalent to a
$p$--adic disk.

Let us note that
$$
{\lambda\over\sqrt{p}}=1
$$
is the maximal possible value of $\lambda$ (the threshold). For
${\lambda\over\sqrt{p}}>1$ any vector (\ref{psi}) has an infinite
norm and therefore does not lie in the Hilbert space. We see that
ultrametricity arise for the maximal value of the order parameter.

Note that if we consider the maximal eigenvalue
${\lambda\over\sqrt{p}}=1$ and take $A={\lambda\over\sqrt{p}}=1$,
then for $Q=A+A^{\dag}$ we obtain $Q=2$. The Wigner semicircle
distribution for $Q$ is supported in the interval $[-2,2]$.  We
see again that the limit $\lambda\to\sqrt{p}-0$ corresponds to the
order parameter $Q$ tending to the maximum.

\section{The $p$--adic representation of the Cuntz algebra}

In the present section we construct  the representation of the
Cuntz algebra in the space $L^2(Z_p)$ of quadratically integrable
functions on a $p$--adic disk. We will call this representation
the $p$--adic representation. Equivalent representations (without
application of $p$--adic analysis) were considered in \cite{BY}.

Let us define the following operators in $L^2(Z_p)$
\be\label{Adagpadic} A^{\dag}_i \xi(x)=\sqrt{p}\theta_1(x-i)
\xi([\frac{1}{p}x]); \ee \be\label{Apadic} A_i
\xi(x)=\frac{1}{\sqrt{p}}\xi(i+px). \ee Here
$$
[x]=x-x(\hbox{mod }1)
$$
for $x\in Q_p$ is the integer part of $x$. $\theta_1(x-i)$ is an
indicator (or characteristic function) of the $p$-adic disk with
the center in  $i$ and the radius $p^{-1}$.

We have the following

\bigskip

\noindent{\bf Theorem 7}\qquad{\sl The operators $A_i^{\dag}$ and
$A_i$ defined by (\ref{Adagpadic}) and (\ref{Apadic}) are mutually
adjoint and satisfy the relations of the Cuntz algebra
(\ref{aac}), (\ref{cuntz}):
$$
A_iA_j^{\dag}=\delta_{ij}.
$$
$$
\sum_{i=0}^{p-1}A_i^{\dag} A_i=1.
$$
}

\bigskip

Let us define the linear functional $\langle\cdot\rangle$ on the
Boltzmann algebra as follows
\begin{equation}\label{padicstate}
\langle A_{I}^{\dag}A_{J} \rangle =p^{-\frac{1}{2}(|I|+|J|)}
\end{equation}
With $A_{I}^{\dag}$ defined by (\ref{Adag}) and $A_{I}$ adjoint.

In the present section we prove that the considered in the
previous section the $p$--adic representation of the Boltzmann
algebra is unitary equivalent to the GNS representation generated
by the state $\langle\cdot\rangle$.

\bigskip

\noindent {\bf Theorem 8}\qquad

{\sl

1) The functional $\langle\cdot\rangle$ is a state.

2) In the corresponding GNS representation the condition
(\ref{cuntz}) is satisfied.

3) The corresponding GNS representation is unitarily equivalent to
the representation realized in the space of (quadratically
integrable) functions on $p$-adic disk by the formula
(\ref{Apadic}):
$$
A_i \xi(x)=\frac{1}{\sqrt{p}}\xi(i+px).
$$
}

\bigskip

We will construct the $p$--adic representation of the Cuntz
algebra by regularization of action of operators from the quantum
Boltzmann algebra on the free coherent states.

Let us introduce some notations. Free coherent state (FCS) $\Phi$
is given by
$$
\Phi=\sum_I \Phi_I \lambda^{|I|} A^{\dag}_I \Omega.
$$
Here $\Omega$ is the vacuum vector in the free Fock space,
$\Phi_I$ satisfies (\ref{cascade}). We will use the notation
$$
\Phi=A^{\dag}_{\Phi} \Omega.
$$

Let us introduce the representation $T$ of the Cuntz algebra in
the space of FCS.

\bigskip

\noindent{\bf Theorem 9}\qquad  {\sl The regularizations
$T^{\dag}_{i}=T(A^{\dag}_{i})$, $T_{i}=T(A_{i})$ of the right
shift operators on FCS
\begin{equation}\label{cuntzcoh}
T^{\dag}_{i}\Phi =\frac{1}{\sqrt{p}} \left(\lambda A^{\dag}_{\Phi}
A^{\dag}_{i} +\Phi_{\emptyset}\right) \Omega=
\frac{1}{\sqrt{p}}\left(\sum_I  \Phi_I \lambda^{|I|+1}
A^{\dag}_{I} A^{\dag}_{i} +\Phi_{\emptyset}\right) \Omega;
\end{equation}
\be\label{cuntzcoh1} T_{i}\Phi={\sqrt{p}}\sum_I  \Phi_{iI}
\lambda^{|I|}A^{\dag}_{I}\Omega= {\sqrt{p}}
\lambda^{-1}\left(A_{i}A^{\dag op}_{\Phi}\right)^{op}\Omega. \ee
where
$$
\left(A^{\dag}_{i_{k-1}}...A^{\dag}_{i_{0}} \right)^{op}=
A^{\dag}_{i_0}...A^{\dag}_{i_{k-1}}.
$$
map the space of FCS into itself and define the representation of
the Cuntz algebra. Moreover, the representation of the Cuntz
algebra defined by the operators $T_{i}$, $T_{i}^{\dag}$ is
unitarily equivalent to the $p$--adic representation.}

\bigskip

\noindent{\bf Theorem 10}\qquad{\sl $p$--Adic representation is
the GNS representation generated by the state on the space of FCS
\begin{equation}\label{afeqtcoh}
\langle X \rangle=(1,X\,1)_{L^2(Z_p)}=\left( \hat 1, AF(X)\, \hat
1 \right)
\end{equation}
Here
$$
\hat 1=\sum_{k=0}^{\infty} \lambda^k \left(\frac{1}{p}
\sum_{i=0}^{p-1}A_i^{\dag}\right)^k \Omega
$$
is the coherent state corresponding to the indicator of $Z_p$ and
$AF$ is the antifock representation. }

\section{Phase transition and coherent states}

In the present paper we consider phase transition in
non--commutative (quantum) system with quantum Boltzmann
statistics.

Let us remind some properties of phase transitions. To describe
phase transitions in quantum systems we use quantum (or algebraic)
probability space --- the algebra of observables ${\cal A}$ with a
state (or expectation) $\langle \cdot \rangle$ on this algebra. In
the Gibbs case this state is equal to \be\label{termalstate}
\langle X \rangle=\hbox{tr }e^{-\beta H}X \ee where $\hbox{tr }$
is a trace on the algebra of observables, $H$ is an element of the
algebra of observables (an Hamiltonian). This state depends on the
{\it control parameter \/} $\beta$ (inverse temperature). For
different physical models we can have different control
parameters. For example, for the models of lasers the typical
control parameter is the intensity of the pumping field.

Using state (\ref{termalstate}) we construct the statistical sum
$$
Z=\langle 1 \rangle=\hbox{tr }e^{-\beta H}
$$
the free energy
$$
F=-\beta^{-1}\ln Z
$$
and other thermodynamical potentials.

If the dependence of thermodynamical potential on $\beta$ is
non--smooth, we say that for such $\beta$ we have phase
transition. For phase transitions we have an effect that the
expectation of certain observables, called the {\it order
parameters\/}, which are equal to zero for the temperature above
the phase transition becomes non--zero after the phase transition.
The typical example is the spontaneous magnetization.

To describe such effects as spontaneous symmetry breaking at the
phase transition N.N.Bo\-go\-liu\-bov introduced the concept of
quasiexpectation \cite{qexp}. We say that the state $\langle \cdot
\rangle_1$ is a quasi\-expectation for the expectation $\langle
\cdot \rangle$ if there exists an operator $H_1$ such that
$$
\langle X \rangle_1=\lim_{\mu\to 0}\lim_{N\to\infty}\hbox{tr
}e^{-\beta (H+\mu H_1)}X
$$
where the limit $N\to\infty$ is the thermodynamic limit. Note that
the limits above do not commute, which makes the definition of
quasiexpectation nontrivial.

Examples of phase transitions in quantum mechanical systems are
the Bose condensation and lasers. For these cases the algebra of
observables ${\cal A}$ is the algebra of canonical commutation
relations, or the CCR algebra (of creation and annihilation
operators) with the relation
$$
[a(k),a^{\dag}(k')]=\delta(k-k')
$$
Let us consider the situation when one of the annihilators (say
$a=a(0)$) is the order parameter and due to the phase transition
the expectation of it becomes non--zero:
$$
\langle a \rangle=\hbox{const}\ne  0
$$
To describe this effect one introduces the coherent states $\psi$
which are the eigenstates of the annihilation operator
$$
a \psi=\lambda \psi
$$
and considers representation of the CCR algebra generated from the
state $\langle\psi, \cdot\psi\rangle$.

In the present paper we extend the outlined above procedure to the
algebra with quantum Boltzmann relations and apply it to describe
the phase transition in disordered system. This will allow us to
develop the approach to phase transitions in disordered systems
which will be an alternative to the replica symmetry breaking
approach.

\section{The noncommutative replica symmetry breaking}

In the standard replica approach the phase transition in
disordered system is described with the help of the replica
symmetry breaking \cite{Parisi}--\cite{SpinGlass}. In the
noncommutative replica approach the phase transition will be
described by noncommutative replica symmetry breaking procedure,
which involves the free coherent states. Since the space of FCS
(free coherent states) and the space of generalized functions on
$p$--adic disk are equivalent, this allows us to derive the
ultrametric space of states which is postulated in the standard
replica aproach.

For the high temperature regime the state of the
Sherrington--Kirkpatrick model is described by the expectations of
large matrices which according to the Wigner theorem give rise to
the vacuum expectation on the quantum Boltzmann algebra in the
free Fock space. The procedure described by the theorems 1 and 2
allows to describe this state equivalently using the Fock state
over the quantum Boltzmann algebra with $p$ degrees of freedom. In
the high temperature regime the expectation of the order parameter
is equal to zero:
$$
\langle A \rangle= 0,\qquad
A={1\over{\sqrt{p}}}\sum_{a=0}^{p-1}A_i
$$
which corresponds to the noncommutative replica symmetry.

After the phase transition in a system with the quantum Boltzmann
statistics, analogously with the Bose condensation and models of
lasers (see the discussion of the previous section), the state
will have the property \be\label{nonzero} \langle A
\rangle=\hbox{const}\ne 0 \ee which we call the noncommutative
replica symmetry breaking condition.

The states satisfying (\ref{nonzero}) are related to the free
coherent states. The space of the free coherent states is
isomorphous to the space of generalized functions over $p$--adic
disk (theorem 6), see \cite{coherent1}, \cite{coherent2},
\cite{coherent3}, and the correspondent representation of the
quantum Boltzmann algebra is realized in $L^2(Z_p)$ (see the
theorems 9 and 10) and is described by theorem  7. This
representation satisfies (\ref{nonzero}) and is not unitarily
equivalent to the Fock representation. Similar representations
were considered by Bratteli, Yorgensen et al \cite{BY} (without
the application of $p$--adic analysis).

Let us note that in the noncommutative replica approach the
operator $Q=A+A^{\dag}$ is the analogue of the Parisi matrix, and
(\ref{nonzero}) corresponds to nonzero expectation of the
Edwards--Anderson order parameter.

Let us formulate the following conjecture, describing the state of
the disordered system after the phase transition.

\bigskip
\noindent {\bf Conjecture 11 }\qquad{\sl The phase transition in
the Sherrington--Kirkpatrick model (and also in a wide class of
disordered models) is described by the state (\ref{padicstate}) on
the quantum Boltzmann algebra generated by annihilators $A_i$,
$i=0,\dots,p-1$. The correspondent representation of the quantum
Boltzmann algebra is given by (\ref{Adagpadic})--(\ref{Apadic}).
The order parameter $Q$ we take to be equal to $Q=A+A^{\dag}$,
where $A$ is the following linear combination of the quantum
Boltzmann annihilators:
$$
A={1\over\sqrt{p}}\sum_{i=0}^{p-1} A_i
$$
}
\bigskip

In the noncommutative replica approach the order parameter $Q$
arise from the limit of large stochastic matrix $J_{ij}$. We did
not introduce any replica matrices.

The ultrametric picture of the state describing the phase
transition in disordered system, which is sometimes discussed as a
result of infinite number of phase transitions, in the
noncommutative replica approach follows from one, but
non--commutative, phase transition (the phase transition is
non--commutative, or quantum, when the order parameter is an
operator).

The introduced conjecture allows us to compute the correlation
functions of the SK model, which are defined as follows:
$$
q^{(1)}={1\over N}\sum_{i}\langle\sigma_i\rangle^2
$$
$$
q^{(2)}={1\over N^2}\sum_{i,j}\langle\sigma_i\sigma_j\rangle^2
$$
and correspondingly
$$
q^{(k)}={1\over N^k}\sum_{i_1\dots
i_k}\langle\sigma_{i_1}\dots\sigma_{i_k}\rangle^2
$$

The standard replica approach gives for the correlators above the
following expression, see \cite{SpinGlass} \be\label{rcorrelator}
q^{(k)}=\lim_{n\to 0}{2\over
n(n-1)}\sum_{a<b}\left[Q_{ab}\right]^{k} \ee where $Q_{ab}$ is the
Parisi matrix.

Introduce the matrix $Q^{(k)}$ with the matrix elements
$$
Q^{(k)}_{ab}=\left[Q_{ab}\right]^{k}
$$
For instance $Q^{(1)}$ is the Parisi matrix.

Formula (\ref{rcorrelator}) means that (up to renormalization and
the limit $n\to 0$), the order parameter is given by
\be\label{averageofQ} q^{(k)}=(\hat 1,Q^{(k)}\hat 1) \ee where
$\hat 1$ is the vector with all the coordinates equal to one.

The noncommutative replica approach suggests the following
conjecture.

\bigskip

\noindent{\bf Conjecture 12}\qquad{\sl The correlator $q^{(k)}$ in
the noncommutative replica approach is approximated by the
following matrix element in the renormalized square product
\be\label{fcsexpression} q^{(k)}=(\hat 1,(cQ)^k \hat 1) \ee where
$\hat 1$ is the free coherent state corresponding to the indicator
of unit $p$--adic disk,
$$
Q=A+A^{\dag},\qquad A={1\over\sqrt{p}}\sum_{i=0}^{p-1}A_i
$$
and $c$ is a constant.}

\bigskip

By (\ref{Adagpadic}), (\ref{Apadic})  expression
(\ref{fcsexpression}) reduces to \be\label{fcsexpression1}
q^{(k)}=(2c)^k\int_{Z_p}d\mu(x)=\left(2c\right)^{k}\mu(V) \ee
where $V$ in the considered case is the unit disk (with the
measure 1). In principle we may consider an equivalent
representation acting on functions on a disk of different measure.
We see that in this approach the correlators of the SK model
reduce to $p$--adic integrals.

\section{Conclusion}

In the present paper we develop the noncommutative replica
approach to phase transitions in disordered systems. Summing up,
this approach looks as follows.

First, by the Wigner theorem, the Sherrington--Kirkpatrick model
in the high temperature regime is described by the quantum
Boltzmann algebra in the free Fock representation. Note that the
high temperature regime here corresponds to the vacuum (zero
temperature) state of the quantum Boltzmann algebra. Then, we
apply the transformation from 1 to $p$ degrees of freedom which is
a morphism of quantum probability spaces. We call this
transformation the {\it Non--commutative Replica Procedure}. In
the noncommutative replica approach we replace the statistic sum
$Z$ of the disordered system by the noncommutative replica
$Z^{(p)}$ of the statistic sum, given by (\ref{Zp}).

Second, we describe phase transition by putting the quantum
Boltzmann algebra into the state with non--zero expectation of the
linear combination of the annihilators $\langle A\rangle\ne 0$,
$A={1\over\sqrt{p}}\sum_{i=0}^{p-1}A_i$. We call the condition
$\langle A\rangle\ne 0$ the condition of breaking of the
noncommutative replica symmetry. To construct such a state we use
the free coherent states and the correspondent $p$--adic
representation of the Cuntz algebra. This representation acts on
functions on the quantum line which is equivalent to $p$--adic
disk.

Third, we conjecture that the $k$--th  correlator of the SK model
is equal to the expectation of the $k$--th degree of the order
parameter $Q=A+A^{\dag}$, $A={1\over\sqrt{p}}\sum_{i=0}^{p-1}A_i$
in the $p$--adic representation. This means that the corresponding
correlators will be given by $p$--adic integrals.

\bigskip

\centerline{\bf Acknowledgements.}

The author would like to thank I.V. Volovich, V.A.Avetisov,
A.H.Bikulov and A.Yu.Khrennikov for discussions and valuable
comments. This work has been partly supported by INTAS (grant No.
9900545), by CRDF grant UM1--2421--KV--02 and The Russian
Foundation for Basic Research (project 02-01-01084).

\end{document}